\documentclass[aps,pra,twocolumn,superscriptaddress,notitlepage,amsmath,amssymb, 10pt]{revtex4-2}

\usepackage[utf8]{inputenc}
\usepackage[T1]{fontenc}
\usepackage{textcomp}
\usepackage{graphicx}
\usepackage{amsfonts}
\usepackage{bm}
\usepackage{braket}
\usepackage{hyperref}
\usepackage{geometry}
\begin{document}

\title{Nonlinear collective-spin dynamics for quantum-enhanced sensing in solid-state platforms}

\author{Le Bin Ho}
\affiliation{Frontier Research Institute for Interdisciplinary Sciences, Tohoku University, Sendai, Japan}
\affiliation{Department of Applied Physics, Graduate School of Engineering, Tohoku University, Sendai, Japan}
\email{ho.bin.le.e3@tohoku.ac.jp}

\begin{abstract}
Collective spin systems in solid-state materials are promising platforms for quantum-enhanced sensing. In systems such as nitrogen-vacancy centers in diamond, donor spins in silicon, and magnetic materials, interactions among many spins can generate collective quantum correlations that improve measurement precision beyond the standard quantum limit. This tutorial develops an intuitive, materials-oriented framework for understanding these effects through nonlinear collective-spin dynamics. We show how nonlinear interactions reshape collective spin states in phase space, redistribute quantum fluctuations, and generate metrologically useful states such as spin-squeezed and entangled states. We then connect these ideal collective dynamics to microscopic interactions and realistic solid-state systems, discussing how interaction strength, disorder, decoherence, and experimental control affect sensing performance. Different platforms and experiments are compared to highlight key material properties, advantages, and limitations. The tutorial focuses on a materials-oriented understanding of nonlinear collective-spin sensing, rather than a comprehensive comparison of experimental platforms or a practical guide to sensing experiments.
\end{abstract}

\maketitle

\section*{Glossary}

\begin{itemize}
\item[] \textbf{Collective spin}: Effective large spin formed by summing individual spin-$1/2$ particles.
\item[] \textbf{Coherent spin state (CSS)}: Product state with all spins aligned along a common direction.
\item[] \textbf{Spin squeezing}: Redistribution of quantum fluctuations that reduces noise in one measurement quadrature below the standard quantum limit.
\item[] \textbf{Standard quantum limit (SQL)}: Precision scaling proportional to $1/\sqrt{N}$ for $N$ independent, uncorrelated particles.
\item[] \textbf{Heisenberg limit}: Ultimate precision scaling proportional to $1/N$, achievable with maximally entangled states.
\item[] \textbf{One-axis twisting (OAT)}: Nonlinear collective interaction proportional to $J_{k}^{2}$, where $k \in \{x,y,z\}$.
\item[] \textbf{Two-axis twisting (TAT)}: Nonlinear collective interaction proportional to $J_{k}^{2} - J_{l}^{2}$, with orthogonal axes $k \neq l$.
\item[] \textbf{Quantum Fisher information (QFI)}: Fundamental measure of the achievable precision in parameter estimation, setting the quantum Cram\'er-Rao bound.
\item[] \textbf{Dephasing}: Loss of phase coherence due to environmental fluctuations or inhomogeneous broadening.
\item[] \textbf{Magnon}: Collective spin-wave excitation in an ordered magnetic material.
\end{itemize}

\section{Introduction}

\begin{figure*}[t]
\centering
\includegraphics[width=0.8\linewidth]{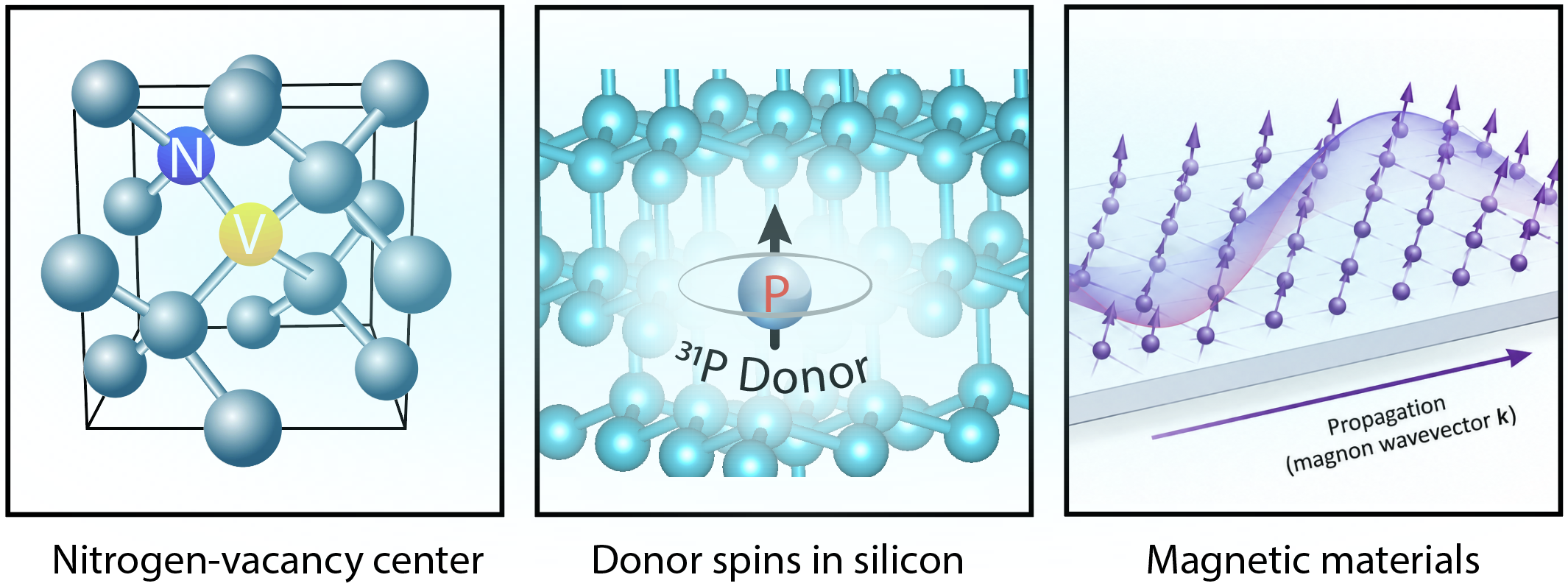}
\caption{Solid-state platforms for collective-spin quantum sensing: nitrogen-vacancy centers in diamond, donor spins in silicon, and collective spin excitations in magnetic materials.}
\label{fig:1}
\end{figure*}

Quantum sensing utilities quantum coherence and entanglement to measure physical parameters with precision beyond classical limits~\cite{ref1}. Solid-state quantum materials provide particularly promising platforms for quantum sensing because they naturally host large ensembles of spins that can be initialized, controlled, and measured collectively. Prominent examples include nitrogen-vacancy (NV) centers in diamond~\cite{ref2,ref3,ref4}, donor spins in silicon~\cite{ref5,ref6,ref7}, and collective spin excitations in magnetic materials~\cite{ref8,ref9,ref10,ref11} as illustrated in Fig.~\ref{fig:1}. These systems combine scalability, robustness, and compatibility with chip-scale architectures, making them attractive candidates for deployable quantum technologies.

For an ensemble of $N$ independent spins, the measurement precision scales as $1/\sqrt{N}$, known as the standard quantum limit (SQL)~\cite{ref12}. This limit arises from uncorrelated quantum fluctuations. In contrast, entanglement introduces correlations among the spins that can reduce these fluctuations and improve the precision toward the Heisenberg scaling of $1/N$~\cite{ref13}. Achieving this enhancement in solid-state systems requires generating entanglement while preserving coherence in noisy environments~\cite{ref14,ref15,ref16}.

Nonlinear interactions among many spins provide a natural way to generate quantum correlations in collective systems. The evolution depends on the collective spin state, so different components of the quantum state accumulate different phases over time. These phase differences gradually deform the collective state in phase space, leading to quantum correlations~\cite{ref17,ref18}.

A simple way to understand this process is to imagine an initially well-aligned spin ensemble, where the quantum uncertainty is distributed symmetrically around the mean spin direction. Under nonlinear evolution, this symmetric distribution gradually becomes distorted and tilted in phase space. Some fluctuations become smaller, while others become larger. This redistribution of quantum fluctuations can reduce measurement noise in certain directions and improve sensing precision beyond the SQL~\cite{ref17,ref18}.

From this viewpoint, the key effect of nonlinear dynamics is the deformation of the collective quantum state in phase space. This deformation creates quantum correlations and transforms simple collective states into useful resources for quantum-enhanced sensing and metrology~\cite{ref18,ref19}.

The primary scope of this tutorial is to connect nonlinear collective-spin dynamics with the material properties and experimental conditions relevant to their realization in solid-state quantum sensors. We focus on how microscopic interactions, disorder, decoherence, and experimental control affect collective dynamics and sensing performance. NV-center ensembles, donor spins in silicon, and magnetic materials are used as major platforms to illustrate materials-dependent opportunities and trade-offs. The experimental discussion focuses on measurable signatures and practical limitations of these dynamics, rather than providing a detailed experimental protocol.

Following this introduction, we first present the collective-spin framework and then explain how nonlinear dynamics generate spin squeezing and other correlated states. We next connect these concepts to solid-state systems by examining the roles of microscopic interactions, disorder, coherence, and other materials-dependent properties. Finally, we discuss their impact on sensing performance, experimental verification, practical limitations, and future directions.

\section{Overview of spin ensemble in solid-state systems}

Many solid-state quantum materials host large numbers of microscopic spins. These spins may originate from point defects in crystals, dopant ions embedded in a host lattice, or collective magnetic excitations in ordered materials. Although each spin is a quantum two-level system, when thousands or millions of them are prepared and controlled together, the ensemble can behave as a single collective object.

\subsection{Collective spin states and the Dicke basis}

A single spin-$1/2$ particle is described by two basis states, $\{\ket{\uparrow},\ket{\downarrow}\}$, defined with respect to a chosen quantization axis. An arbitrary pure state of the spin can be written as
\begin{equation}
\ket{\psi} = \alpha \ket{\uparrow} + \beta \ket{\downarrow},
\label{eq:1}
\end{equation}
where the complex amplitudes satisfy $|\alpha|^{2} + |\beta|^{2} = 1$. This state corresponds to a point on the Bloch sphere.

When considering an ensemble of $N$ spin-$1/2$ particles, the full system formally resides in a Hilbert space of dimension $2^{N}$, which grows exponentially with the number of spins. Many quantum-sensing protocols, however, rely on collective control in which all spins are addressed identically~\cite{ref20}. In this case, it is convenient to introduce the collective spin operators
\begin{equation}
J_{\alpha} = \frac{1}{2}\sum_{i=1}^{N}\sigma_{\alpha}^{(i)}, \quad \alpha \in \{x,y,z\},
\label{eq:2}
\end{equation}
where $\sigma_{\alpha}^{(i)}$ denotes the Pauli operator acting on the $i$-th spin. These operators represent the total angular momentum of the ensemble and satisfy the usual angular-momentum commutation relations, $[J_{x},J_{y}] = iJ_{z}$, together with cyclic permutations. For convenience, we also define the collective spin vector $\mathbf{J} = (J_{x},J_{y},J_{z})$.

\begin{figure}[t]
\centering
\includegraphics[width=\columnwidth]{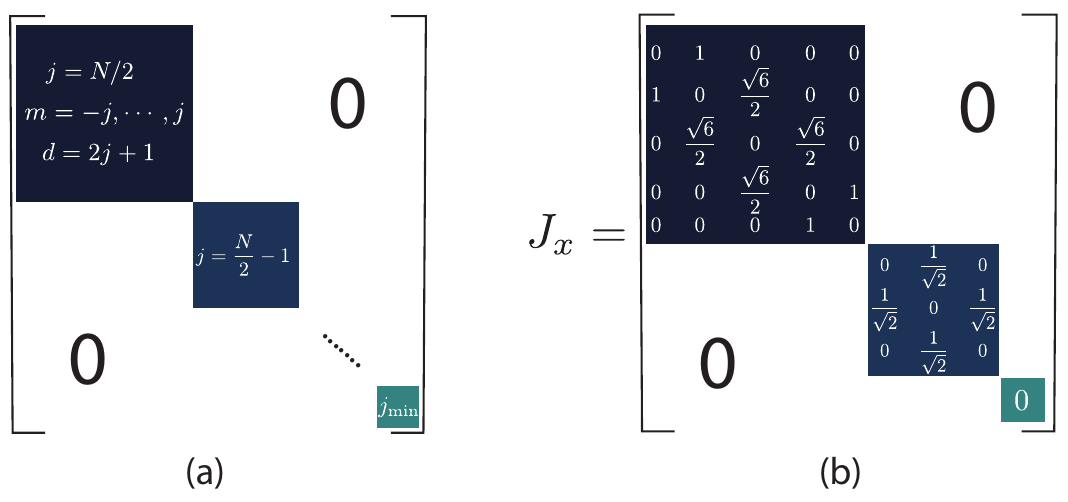}
\caption{Block-diagonal structure of the collective spin operators in the Dicke basis. (a) Each block corresponds to a fixed total-spin sector $j$. (b) Example for $N=4$, where the Dicke-space dimension $d_D=9$ is much smaller than the full Hilbert-space dimension $2^4=16$.}
\label{fig:2}
\end{figure}

The eigenstates of the collective operators $J^{2}$ and $J_{z}$ define the Dicke basis, denoted by $\ket{j,m}$, which satisfies
\begin{equation}
J^{2}\ket{j,m} = j(j+1)\ket{j,m}, \quad J_{z}\ket{j,m} = m\ket{j,m}.
\label{eq:3}
\end{equation}
Here, $j$ is the total spin quantum number, and $m = -j,-j+1,\ldots,j$ are the spin projections along the $z$-axis. For an ensemble of $N$ spin-$1/2$ particles, the allowed total-spin values are
\begin{equation}
j = \left(\frac{N}{2},\ \frac{N}{2}-1,\ \frac{N}{2}-2,\ \ldots,\ j_{\min}\right),
\label{eq:4}
\end{equation}
where $j_{\min} = 0$ for even $N$ and $j_{\min} = 1/2$ for odd $N$. As illustrated in Fig.~\ref{fig:2}(a), the collective operators exhibit a block-diagonal structure, with each block corresponding to a fixed total-spin sector $j$. Each sector forms a subspace of dimension $d_{j} = 2j+1$. As a result, the effective Hilbert-space dimension is dramatically reduced from the full $2^{N}$-dimensional space to the Dicke-space dimension~\cite{ref21}
\begin{equation}
d_{D} =
\begin{cases}
\dfrac{(N+2)^{2}}{4}, & \text{for even } N, \\[6pt]
\dfrac{(N+1)(N+3)}{4}, & \text{for odd } N.
\end{cases}
\label{eq:5}
\end{equation}
For example, when $N=4$, the allowed spin sectors are $j = 2,1,0$, with corresponding subspace dimensions $5 \oplus 3 \oplus 1$. The Dicke-space dimension is therefore $d_{D} = 5+3+1 = 9$, which is much smaller than the full Hilbert-space dimension $2^{4} = 16$, as illustrated in Fig.~\ref{fig:2}(b).

For a permutation-symmetric initial state and collective dynamics that preserve this symmetry, the evolution remains within the maximum-spin sector $J = N/2$. In this case, the Dicke states $\ket{j,m}$ correspond to symmetric superpositions of configurations with $j+m$ spins in the state $\ket{\uparrow}$ and $j-m$ spins in the state $\ket{\downarrow}$. This collective description simplifies the physical picture: rather than dealing with an exponentially large Hilbert space, the ensemble can be visualized as a single large spin represented by a vector on a sphere.

However, in solid-state spin ensembles, spatial disorder, inhomogeneous fields, strain, and variations in spin-spin coupling can break this symmetry by exposing individual spins to different local environments and interactions. The resulting inhomogeneity drives the dynamics away from the fully symmetric collective-spin sector and can suppress the many-body correlations. Controlling these effects is therefore a central challenge for realizing collective-spin sensing in solid-state systems~\cite{ref22}.

\begin{figure}[b]
\centering
\includegraphics[width=0.7\columnwidth]{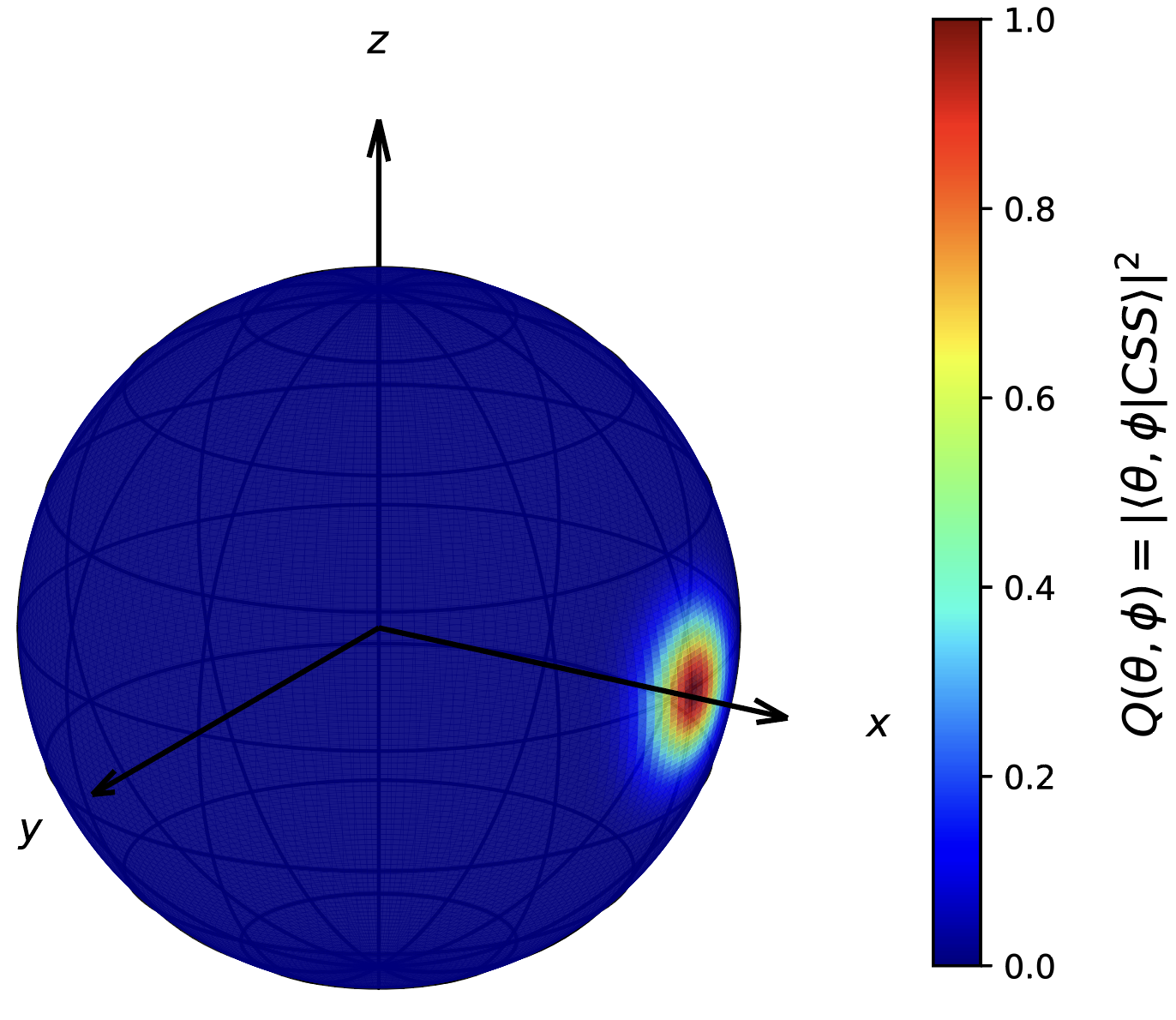}
\caption{Husimi $Q$ distribution of a coherent spin state (CSS) on the Bloch sphere. The circular, isotropic distribution reflects the uncorrelated quantum projection noise that sets the standard quantum limit.}
\label{fig:3}
\end{figure}

\subsection{Coherent spin states}

A particularly important state for sensing is the coherent spin state (CSS)~\cite{ref18}. This state is obtained when all spins are aligned along the same direction. For example, if every spin points along the $x$-axis, the collective state can be written as
\begin{equation}
\ket{\mathrm{CSS}} = \left(\frac{\ket{\uparrow}+\ket{\downarrow}}{\sqrt{2}}\right)^{\otimes N}.
\label{eq:6}
\end{equation}
In the Dicke basis, the same state can be expressed as
\begin{equation}
\ket{\mathrm{CSS}} = \sum_{m=-\frac{N}{2}}^{\frac{N}{2}} \sqrt{\binom{N}{\frac{N}{2}+m}}\left(\frac{1}{2}\right)^{\frac{N}{2}} \left|\frac{N}{2},m\right\rangle.
\label{eq:7}
\end{equation}
This expression shows that a CSS is a binomial superposition of Dicke states with different spin projections $m$. When the mean spin points along the $x$-direction, the distribution is centered at $m=0$. The expectation value of the collective spin is $\langle J_{x}\rangle = \frac{N}{2}$, while the fluctuations perpendicular to the mean spin direction satisfy $(\Delta J_{y})^{2} = (\Delta J_{z})^{2} = \frac{N}{4}$, where $\Delta O = \sqrt{\langle O^{2}\rangle - \langle O\rangle^{2}}$.

The Husimi $Q$ function provides a positive phase-space representation of the collective-spin state. It is defined by $Q(\theta,\phi) \propto \langle \theta,\phi|\rho|\theta,\phi\rangle$, where $\ket{\theta,\phi}$ is a CSS oriented along $(\theta,\phi)$. For a pure state, $Q(\theta,\phi)$ is proportional to the squared overlap with this coherent state. The distribution provides a geometric representation of the collective state on the Bloch sphere: regions of large $Q$ indicate the dominant spin orientations, while the width and shape of the distribution characterize the collective fluctuations.

For a CSS, the Husimi distribution forms a circular region centered on the mean spin direction, as shown in Fig.~\ref{fig:3}. This isotropic width reflects the quantum projection noise of an uncorrelated spin ensemble and sets the SQL for sensing with unentangled states.

\subsection{Nonlinear collective-spin dynamics}

Once a spin ensemble is described in terms of collective spin operators, the next key question is how to generate correlations between spins. Such correlations are absent in a CSS, where all spins evolve independently. Instead, they arise from interactions that couple the spins collectively, giving rise to nonlinear collective-spin dynamics~\cite{ref23,ref24}.

\begin{figure}[t]
\centering
\includegraphics[width=\columnwidth]{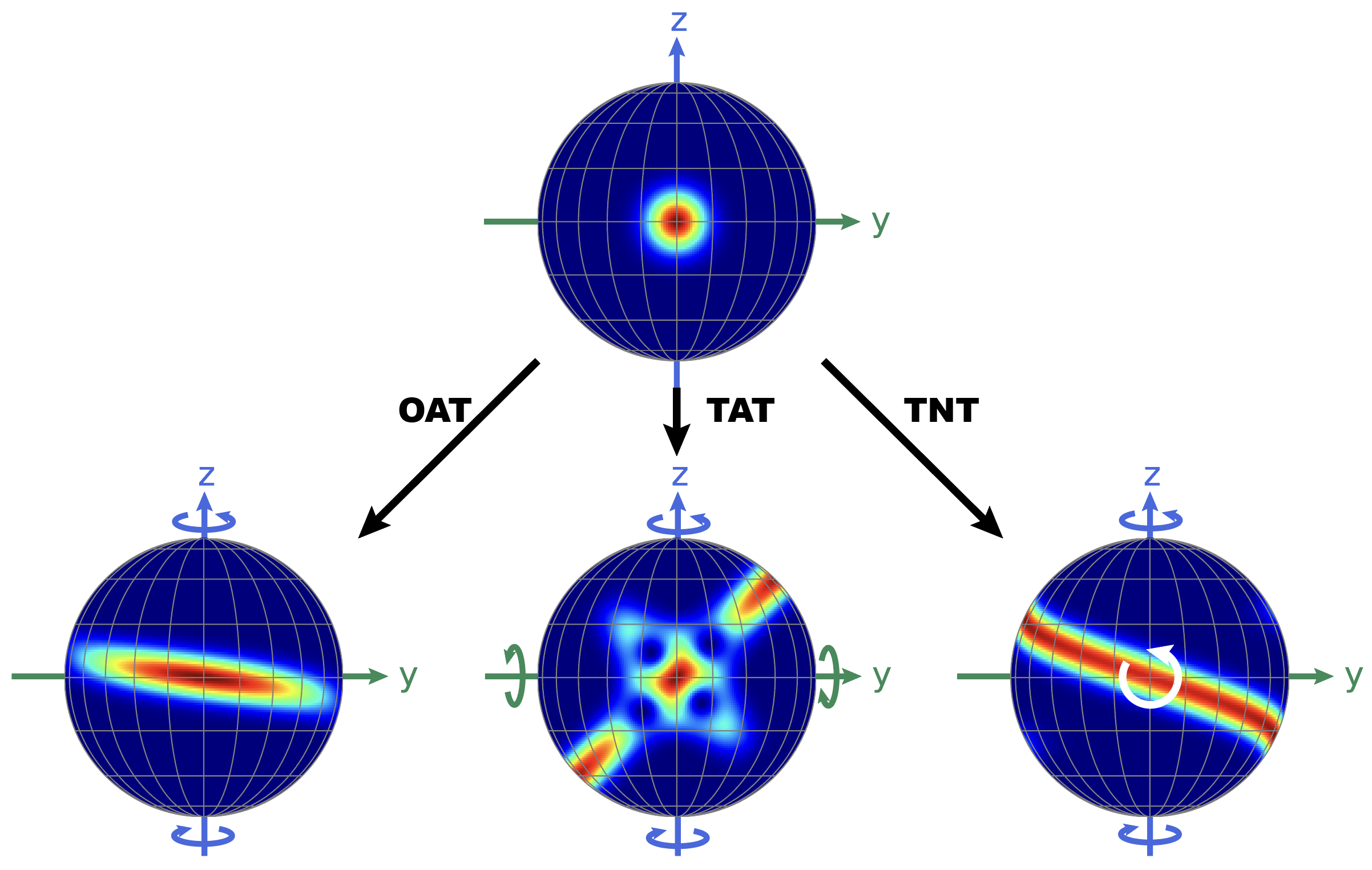}
\caption{Nonlinear deformation of the Husimi $Q$ distribution in phase space under one-axis twisting. The initially circular CSS distribution shears into an ellipse (spin squeezing) and, at longer times, splits into multiple components.}
\label{fig:4}
\end{figure}

A broad class of these interactions can be described by Hamiltonians that depend quadratically on collective spin operators. A paradigmatic example is the one-axis twisting (OAT) Hamiltonian~\cite{ref17,ref25},
\begin{equation}
H_{\mathrm{OAT}} = \chi J_{z}^{2},
\label{eq:8}
\end{equation}
where $\chi$ characterizes the effective collective interaction strength. Under ideal OAT evolution, each spin-projection component acquires a phase proportional to $m^{2}$, where $m$ is the eigenvalue of $J_{z}$. This projection-dependent phase accumulation drives the nonlinear evolution of the collective state in phase space, as illustrated in Fig.~\ref{fig:4}.

The ideal OAT assumes a uniform collective nonlinear interaction. In solid-state defect ensembles, however, the spins may occupy nonuniform or randomly distributed positions within the host lattice. For dipolar interactions, the coupling between spins $i$ and $j$ depends strongly on their separation and relative orientation,
\begin{equation}
J_{ij} \propto \frac{1-3\cos^{2}\theta_{ij}}{r_{ij}^{3}},
\label{eq:9}
\end{equation}
where $r_{ij}$ is the interspin distance and $\theta_{ij}$ is the angle between the spins vector and the quantization axis. Spatial disorder therefore produces a distribution of $J_{ij}$, resulting in nonuniform spin-spin interactions that depart from the collective coupling assumed in the ideal OAT model. Realizing OAT dynamics in such systems thus requires sufficiently uniform effective interactions or suitable Hamiltonian-engineering techniques~\cite{ref22}.

Another important interaction is the two-axis twisting (TAT) Hamiltonian~\cite{ref26},
\begin{equation}
H_{\mathrm{TAT}} = \chi\left(J_{x}^{2}-J_{y}^{2}\right),
\label{eq:10}
\end{equation}
which redistributes quantum fluctuations more rapidly and can generate stronger correlations on shorter timescales than the OAT. As a result, TAT dynamics are effective for producing highly correlated states within experimentally relevant coherence times.

A related and experimentally important class of nonlinear dynamics is the twist-and-turn (TNT) scheme, which combines collective twisting with a linear rotation~\cite{ref27}. A minimal model is described by the Hamiltonian
\begin{equation}
H_{\mathrm{TNT}} = \chi J_{z}^{2} + \Omega J_{x},
\label{eq:11}
\end{equation}
where $\Omega$ is the rotation rate about an axis orthogonal to the twisting axis. For $\Omega=0$, the Hamiltonian reduces to OAT, while a finite $\Omega$ continuously rotates the collective spin during the nonlinear evolution. This combined dynamics accelerates the redistribution of quantum fluctuations and can generate strong squeezing faster than OAT alone. The shorter evolution time reduces exposure to decoherence, making TNT attractive for systems with limited coherence times.

The physical effect of these nonlinear interactions can be understood geometrically. For an initial CSS, the uncertainty distribution in the plane perpendicular to the mean spin direction is circular. Under nonlinear evolution, different parts of the collective wavefunction accumulate different phases. This causes the initially symmetric distribution to become distorted and sheared in phase space. At early evolution times, this shearing produces an anisotropic uncertainty distribution. At later times, the deformation can become strong enough to split the wavefunction into multiple components, leading to macroscopic quantum superposition states.

\subsection{Spin squeezing states}
\label{sec:squeezing}

\begin{figure}[t]
\centering
\includegraphics[width=\columnwidth]{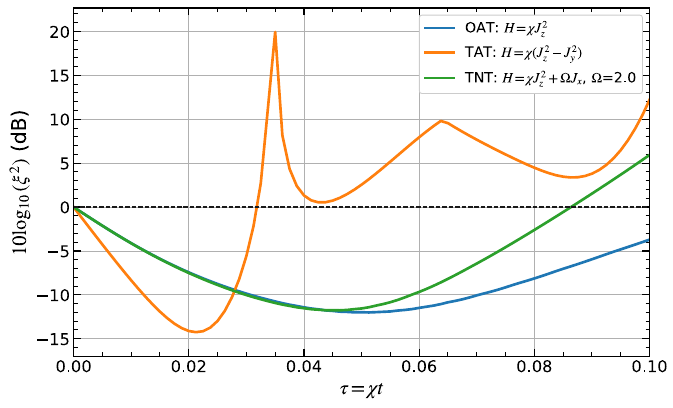}
\caption{Wineland squeezing parameter $\xi^2$ as a function of the scaled evolution time $\chi t$ for OAT, TAT, and TNT dynamics.}
\label{fig:5}
\end{figure}

One of the most important states generated by nonlinear collective dynamics is the spin-squeezed state. A CSS does not contain correlations between individual spins, and its measurement precision is therefore limited by the SQL. In a spin-squeezed state, the transverse quantum fluctuations become anisotropic. The uncertainty is reduced along one direction and increased along the conjugate direction, consistent with the uncertainty principle. This deformation provides a simple geometric picture of spin squeezing. The reduced uncertainty improves sensitivity to small rotations when the measurement axis is aligned with the squeezed direction. A widely used measure of spin squeezing is the Wineland parameter~\cite{ref28},
\begin{equation}
\xi^{2} = \frac{N\,(\Delta J_{\bot})^{2}}{|\langle \mathbf{J}\rangle|^{2}},
\label{eq:12}
\end{equation}
where $J_{\bot}$ denotes the collective spin component orthogonal to the mean spin direction. When $\xi^{2}<1$, the state is spin squeezed and enables measurement precision beyond the SQL. In Fig.~\ref{fig:5}, we show the squeezing Wineland parameter for different collective spin dynamics OAT, TAT, and TNT.

\begin{figure*}[t]
\centering
\includegraphics[width=0.8\linewidth]{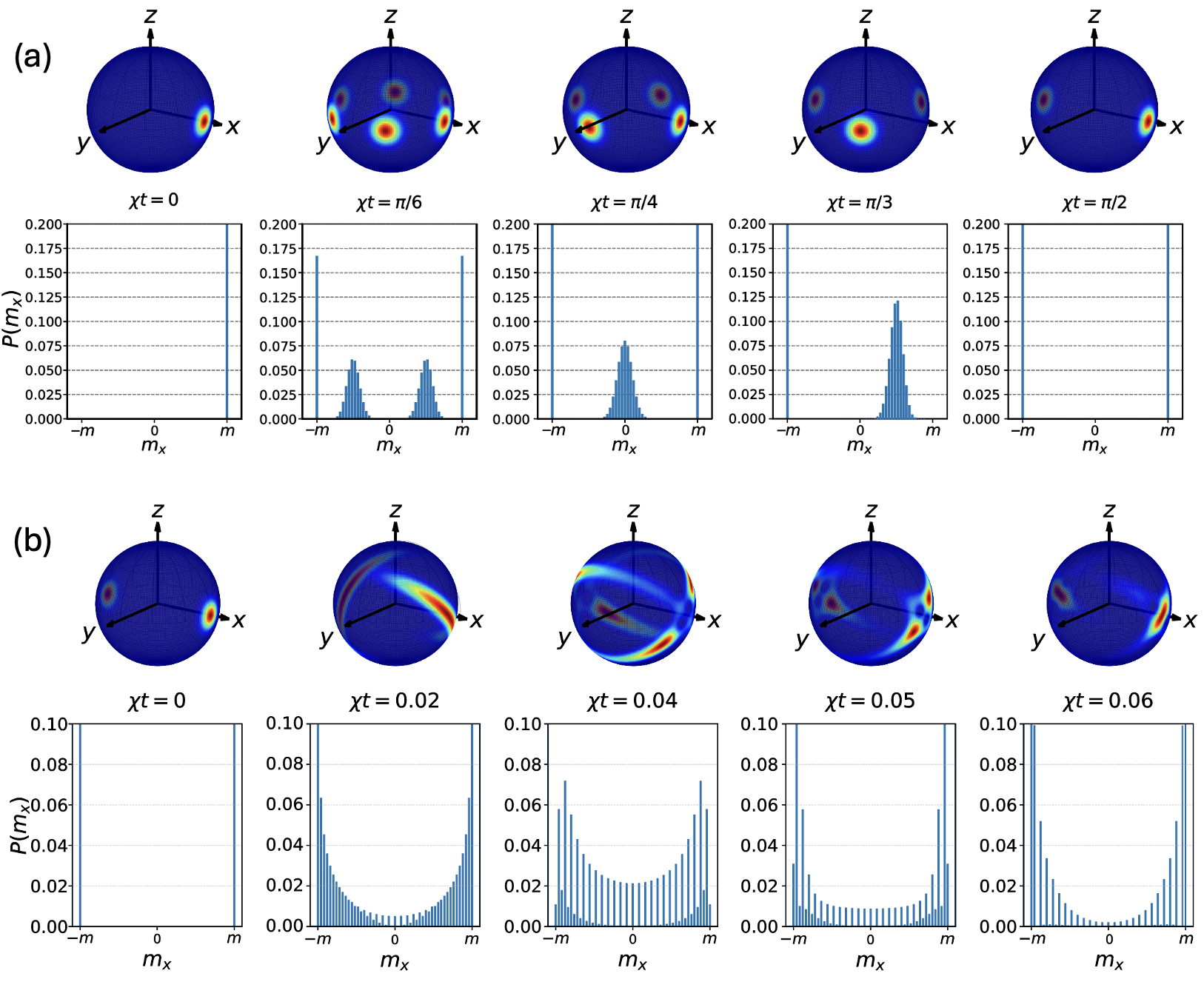}
\caption{(a) Interference dynamics from a CSS to the GHZ-like superposition state of Eq.~\eqref{eq:14} generated by OAT. (b) Continued dynamics: the GHZ-like state generated by OAT at $\chi t=\pi/2$ serves as the initial state for subsequent TAT dynamics in the $y$-$z$ plane. Upper panels: redistribution and refocusing of the two peaks in the Husimi $Q$ distribution. Lower panels: corresponding projective probabilities $P(m_x)$~\cite{ref17}.}
\label{fig:6}
\end{figure*}

On the Husimi phase-space representation, this process corresponds to a deformation of the uncertainty region associated with the CSS (see illustration in Fig.~\ref{fig:4}). The initially circular distribution becomes elliptical, with the short axis indicating the squeezed direction. Aligning the sensing rotation with this axis allows the ensemble to detect small phase shifts with enhanced precision. Spin squeezing therefore represents the first stage of quantum enhancement generated by nonlinear collective-spin dynamics, linking the microscopic interaction Hamiltonian directly to improved sensing performance.

Spin squeezing also provides a quantitative witness of multipartite entanglement. S{\o}rensen and M{\o}lmer established a relation between the transverse spin variance and the mean collective spin that can be used to detect multipartite entanglement~\cite{ref24}. More recently, Block \textit{et al.} showed that short-range interactions in finite-temperature easy-plane ferromagnets can generate scalable spin squeezing~\cite{ref29}. They also identified a transition between scalable and nonscalable squeezing near the equilibrium ordering boundary. These results show that scalable squeezing does not require ideal all-to-all interactions and can also arise in locally interacting spin systems.

\subsection{Interference perspective and GHZ state generation}

Nonlinear collective-spin dynamics can generate highly entangled states when the evolution time becomes sufficiently long. A particularly important example arises from the OAT Hamiltonian, $H_{\mathrm{OAT}} = \chi J_{z}^{2}$, where $\chi$ denotes the interaction strength. The corresponding time-evolution operator is $U(t) = e^{-i\chi t J_{z}^{2}}$. To understand the resulting dynamics, it is convenient to expand an initial CSS polarized along the $x$-direction in the Dicke basis, $\ket{\mathrm{CSS}} = \sum_{m=-j}^{j} c_{m}\ket{j,m}$, where $j=N/2$ and the coefficients $c_{m}$ form a binomial distribution centered at $m=0$. Under the OAT evolution, each Dicke component acquires a quadratic, projection-dependent phase, $\ket{j,m} \longrightarrow e^{-i\chi t m^{2}}\ket{j,m}$, so that the state evolves as
\begin{equation}
\ket{\psi(t)} = \sum_{m=-j}^{j} c_{m}\, e^{-i\chi t m^{2}} \ket{j,m}.
\label{eq:13}
\end{equation}
Because the accumulated phase depends on $m^{2}$, different spin-projection sectors evolve at different rates. This projection-dependent phase accumulation generates quantum interference among the Dicke components of the many-body wavefunction. A particularly striking situation occurs at the evolution time $t=\pi/(2\chi)$. At this time, the phase factor becomes $e^{-i\chi t m^{2}} = e^{-i\frac{\pi}{2}m^{2}}$, and since $m^{2}$ has different parity for even and odd $m$, the evolution effectively separates the Dicke components into two groups. Using the identity $e^{-i\pi J_{z}}\ket{j,m} = (-1)^{m}\ket{j,m}$, the evolved state can be written as a coherent superposition of two collective spin states pointing in opposite directions,
\begin{equation}
\ket{\psi(\pi/2\chi)} \propto \ket{\mathrm{CSS}_{+x}} + i\ket{\mathrm{CSS}_{-x}}.
\label{eq:14}
\end{equation}
These two components correspond to collective spins located at opposite poles of the sphere along the $x$-axis. The resulting state represents a macroscopic superposition of two distinct collective spin orientations. Figure~\ref{fig:6}(a) illustrates the interference dynamics from a CSS state to the state in Eq.~\eqref{eq:14}. Up to a global rotation, this state is equivalent to the Greenberger-Horne-Zeilinger (GHZ) state,
\begin{equation}
\ket{\mathrm{GHZ}} = \frac{1}{\sqrt{2}}\left(\ket{\uparrow\uparrow\cdots\uparrow} + \ket{\downarrow\downarrow\cdots\downarrow}\right),
\label{eq:15}
\end{equation}
which exhibits maximal quantum correlations among all spins. In this regime, the collective state has a variance of order $N^{2}$ in the generator $J_{x}$, leading to Heisenberg-limited phase sensitivity scaling as $1/N$.

From an interference perspective, nonlinear evolution, such as that generated by $\exp(-i\chi t J_{z}^{2})$, imprints projection-dependent phases that reshape the many-body wavefunction in collective phase space. At intermediate evolution times, this process first produces spin squeezing, while at longer times the interference becomes strong enough to split the wavefunction into multiple components. This ultimately leads to macroscopic superposition states, such as Schr\"odinger-cat or GHZ states, formed through quantum self-interference.

Figure~\ref{fig:6}(b) continues the dynamics shown in Fig.~\ref{fig:6}(a). The GHZ-like state generated by OAT at $\chi t = \pi/2$ serves as the initial state for subsequent TAT dynamics in the $y$-$z$ plane. The upper panels show the redistribution and refocusing of the two peaks in the Husimi $Q$ distribution, while the lower panels show the corresponding projective probabilities $P(m_x)$~\cite{ref17}.

In solid-state spin ensembles, the physical realization of such nonlinear collective interactions depends strongly on the material platform. In two-dimensional NV ensembles, the secular magnetic dipole-dipole interaction contains both exchange and Ising components. By choosing the quantization axis appropriately, their ensemble-averaged difference can produce OAT-like dynamics, while Floquet pulse sequences can engineer TAT-like evolution~\cite{ref22,ref30}. Optical AC Stark shifts, by contrast, are generally linear in $J_{z}$ and do not directly generate squeezing. An effective $J_{z}^{2}$ interaction can emerge when the light shift depends on the collective spin state, as in dispersive cavity-feedback schemes~\cite{ref26}. In magnetic materials, crystalline anisotropy and magnon-magnon interactions can generate a magnon Kerr nonlinearity, which corresponds to a quadratic collective-spin interaction~\cite{ref13}. For donor spins in silicon, dipolar and exchange interactions are generally nonuniform and depend on the donor configuration. Existing experiments demonstrate coherent spin control and long coherence times, but have not yet realized collective OAT squeezing~\cite{ref5,ref6,ref7}.

\section{Application to quantum sensing}

The purpose of engineering nonlinear collective-spin dynamics is to improve the precision with which an external parameter can be estimated. In a typical sensing protocol, a spin ensemble is prepared in an initial state, allowed to interact with a parameter $\theta$ (for example, a magnetic field $B$), and then measured collectively.

For many sensing platforms, the parameter couples to the ensemble through a Hamiltonian of the form
\begin{equation}
H = \gamma B J_{z}, \qquad U(t) = e^{-i\gamma B t J_{z}},
\label{eq:16}
\end{equation}
where $B$ is the magnetic field and $\gamma$ is the gyromagnetic ratio. During an interaction time $t$, the ensemble undergoes a collective rotation about the $z$-axis,
\begin{equation}
\ket{\psi(\theta)} = e^{-i\gamma B t J_{z}}\ket{\psi_{0}} = e^{-i\theta J_{z}}\ket{\psi_{0}},
\label{eq:17}
\end{equation}
where $\ket{\psi_{0}}$ is the initial state and $\theta = \gamma B t$. The sensing task is therefore to estimate the rotation angle $\theta$ with the smallest possible uncertainty.

\subsection{Precision limits and the standard quantum limit}

A convenient way to estimate the phase is to measure a transverse collective spin component after the rotation. For small rotations around the $z$-axis, the uncertainty in estimating $\theta$ follows from the error-propagation formula
\begin{equation}
\Delta\theta = \frac{\Delta J_{\bot}}{|\partial_{\theta}\langle J_{\bot}\rangle|},
\label{eq:18}
\end{equation}
where $J_{\bot}$ denotes a collective spin component transverse to the mean spin direction.

Concretely, consider an initial state $\ket{\psi_{0}}$ polarized along the $x$-direction, so that the mean collective spin points along $x$, $\langle\mathbf{J}\rangle = (\langle J_{x}\rangle,\,0,\,0)$. The transverse plane is then the $y$-$z$ plane, and any transverse operator can be written as $J_{\bot}(\phi) = J_{y}\cos\phi + J_{z}\sin\phi$. For a CSS, the transverse fluctuations are isotropic, so all directions perpendicular to the mean spin are equivalent. In particular, $\Delta J_{y} = \sqrt{N/4}$, $\partial_{\theta}\langle J_{\bot}\rangle = \langle J_{x}\rangle = N/2$. This leads to
\begin{equation}
\Delta\theta_{\mathrm{SQL}} = \frac{\sqrt{N/4}}{N/2} = \frac{1}{\sqrt{N}},
\label{eq:19}
\end{equation}
which is the SQL. This limit reflects the independent quantum projection noise of individual spins.

\subsection{Spin squeezing and redistribution of quantum fluctuations}

Spin squeezing improves sensing precision by reducing the fluctuation of the collective-spin component used for phase estimation. Using the squeezing parameter introduced in Sec.~\ref{sec:squeezing}, the transverse fluctuation can be written as $\Delta J_{\bot} = \xi\sqrt{N/4}$, while the mean collective spin remains approximately $|\langle\mathbf{J}\rangle| \simeq N/2$. Applying the same error-propagation formula yields
\begin{equation}
\Delta\theta = \frac{\Delta J_{\bot}}{|\langle\mathbf{J}\rangle|} = \frac{\xi}{\sqrt{N}}.
\label{eq:20}
\end{equation}
Spin squeezing occurs when $\xi<1$, demonstrating that squeezing improves metrological precision by reducing the relevant transverse quantum fluctuation below that of a CSS.

Geometrically, nonlinear evolution, such as that generated by $H=\chi J_{z}^{2}$, shears the collective quantum state in phase space. The initially circular uncertainty region associated with a CSS becomes elliptical. When the sensing rotation is aligned with the squeezed axis of this ellipse, the reduced width directly translates into a smaller estimation error.

\subsection{Quantum Fisher information and ultimate precision bounds}

A more general description of sensing precision is provided by the quantum Cram\'er-Rao bound~\cite{ref31},
\begin{equation}
\Delta\theta \geq \frac{1}{\sqrt{F_{Q}}},
\label{eq:21}
\end{equation}
where $F_{Q}$ is the quantum Fisher information (QFI). For a pure state $\ket{\psi(\theta)}$, the QFI is given by~\cite{ref31}
\begin{equation}
F_{Q} = 4\left(\langle \partial_{\theta}\psi(\theta)|\partial_{\theta}\psi(\theta)\rangle - \left|\langle\psi(\theta)|\partial_{\theta}\psi(\theta)\rangle\right|^{2}\right).
\label{eq:22}
\end{equation}
Now suppose the parameter $\theta$ is encoded through a unitary evolution generated by $J_{z}$. In this case, the quantum Fisher information reduces to $F_{Q} = 4(\Delta J_{z})^{2}$. For a CSS, one finds $F_{Q}=N$, which again leads to the SQL scaling. By contrast, maximally entangled states such as GHZ states achieve $F_{Q}=N^{2}$, corresponding to the Heisenberg limit $\Delta\theta_{\mathrm{HL}} = 1/N$.

In practice, nonlinear collective-spin dynamics generate intermediate states that lie between these two extremes~\cite{ref23,ref24}. Spin-squeezed states can provide a useful compromise between metrological enhancement and experimental robustness, offering improved precision while generally remaining less sensitive to decoherence than maximally entangled states.

\section{Experimental realization and verification}

In experiments, quantum enhancement is not measured directly but inferred from collective-spin signals and their fluctuations. A typical experiment prepares the same initial state, applies the nonlinear dynamics and sensing sequence, and measures a collective-spin observable. Repeating this procedure provides the statistics needed to quantify squeezing, signal amplification, and sensing precision.

Experimental demonstrations have progressed from atomic ensembles to controllable many-body systems and solid-state spin ensembles. An early experiment transferred nonclassical optical fluctuations to an ensemble of approximately $10^{7}$ cold atoms and observed reduced collective-spin noise~\cite{ref32}. More recently, Bornet \textit{et al.} generated spin squeezing through dipolar XY interactions in programmable Rydberg arrays of up to 100 atoms~\cite{ref33}. They measured $-3.5\pm0.3$~dB of squeezing before detection-error correction and approximately $-5\pm0.3$~dB after correction.

Room-temperature NV-center experiments have extended these ideas to solid-state ensembles. Wu \textit{et al.} reduced the effects of positional disorder by selecting a more ordered subensemble and measured $-0.50\pm0.13$~dB of spin squeezing~\cite{ref22}. Gao \textit{et al.} used native dipolar interactions and Floquet control to engineer OAT- and TAT-like dynamics in a quasi-two-dimensional NV layer and observed $6.7(6)\%$ signal amplification with an asymmetric many-body echo~\cite{ref30}. The former directly measures reduced spin fluctuations, whereas the latter demonstrates interaction-based amplification of a sensing signal.

\begin{figure}[t]
\centering
\includegraphics[width=\columnwidth]{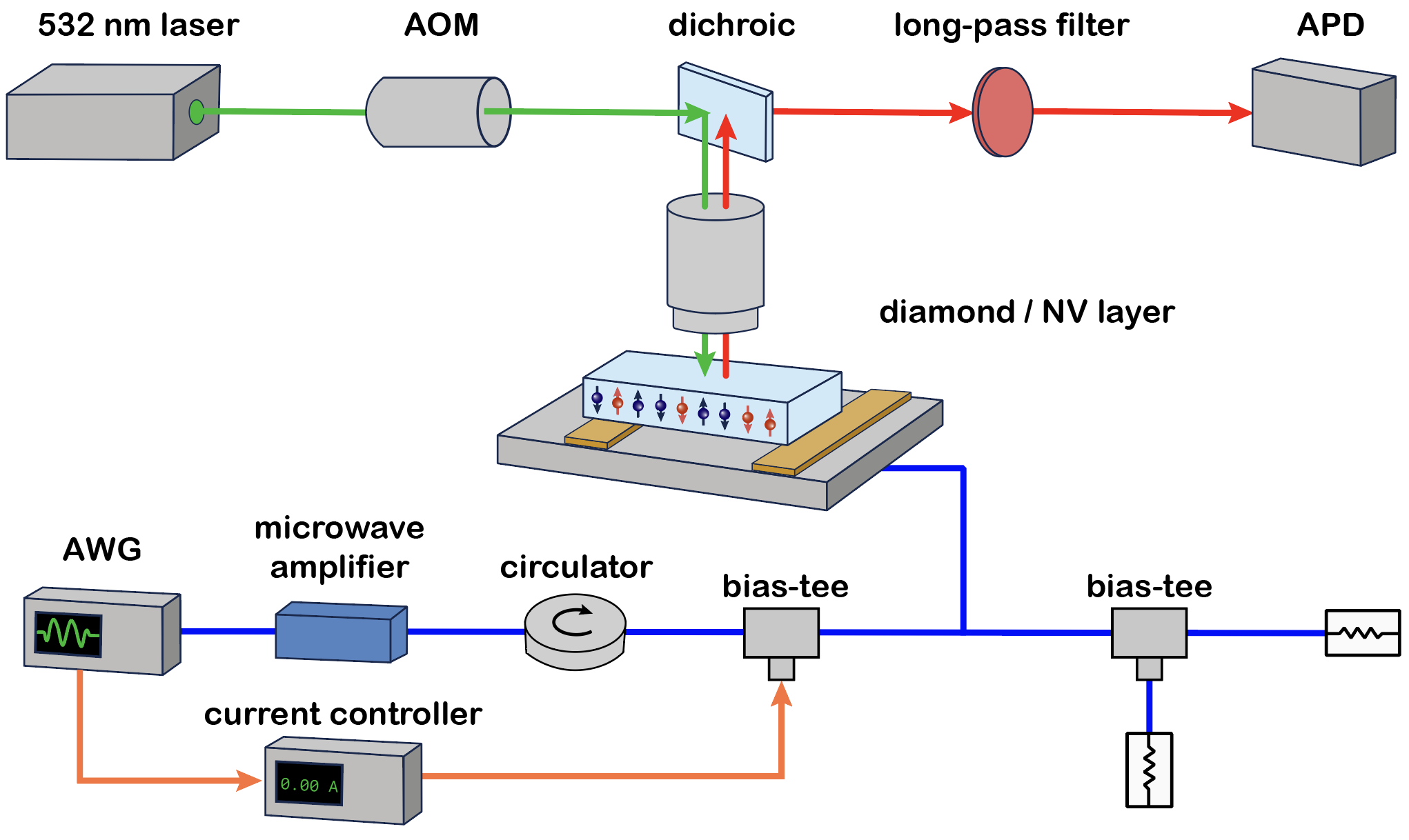}
\caption{Experimental setup for a room-temperature NV-ensemble solid-state sensor, following Gao \textit{et al.}~\cite{ref30}. A synchronized arbitrary waveform generator controls the microwave pulses and the pulsed sensing-field current; optical initialization and fluorescence readout are performed via a 532-nm confocal beam and an avalanche photodiode.}
\label{fig:7}
\end{figure}

In NV ensembles, optical excitation initializes the spins, microwave pulses control the collective dynamics, and fluorescence provides the readout. Figure~\ref{fig:7} illustrates the experimental setup used by Gao \textit{et al.}~\cite{ref30}. A synchronized arbitrary waveform generator controls the microwave pulses and the pulsed current that generates the sensing field. The microwave signal is delivered through a coplanar waveguide, while a 532-nm confocal beam provides optical initialization and the resulting fluorescence is detected with an avalanche photodiode.

The scale of a solid-state ensemble should be specified by both the spin density and the addressed volume or area. For a thin NV layer, ppm$\cdot$nm describes an areal density obtained by integrating the concentration over the layer thickness. Wu \textit{et al.} reported approximately 8~ppm$\cdot$nm for the interacting NV ensemble~\cite{ref22}, while Gao \textit{et al.} reported 19~ppm$\cdot$nm for each NV orientation in a 9-nm-thick layer~\cite{ref30}. The actual number of participating spins also depends on the addressed area, spin polarization, and selection protocol.

To establish quantum enhancement, experimental noise must be separated from the intrinsic spin fluctuations. Detector noise, control errors, magnetic-field drift, and finite sampling can modify both the measured signal and its variance. The readout and technical noise must therefore be calibrated, and the measured sensitivity should be compared with the SQL under the same experimental conditions.

\section{Materials-dependent limitations and comparison}

The realization of nonlinear collective-spin sensing depends on the balance between interaction strength, coherence, and disorder. Stronger spin-spin interactions can generate correlations more rapidly, while decoherence and interaction inhomogeneity reduce the lifetime and uniformity of these correlations. These effects differ substantially between solid-state platforms.

\begin{table*}[t]
\caption{Experimentally reported conditions, nonlinear observables, and coherence or damping scales for selected solid-state spin platforms. Nonlinear observables are listed only when experimentally demonstrated; linear coupling rates and coherence times are not interpreted as effective twisting strengths.}
\label{tab:1}
\begin{ruledtabular}
\small
\begin{tabular}{p{2.0cm}p{2.9cm}p{3.5cm}p{3.0cm}p{2.7cm}}
Platform & Conditions & Nonlinear observable & Coherence/\allowbreak damping & Interpretation \\
\colrule
2D NV ensembles in diamond \cite{ref22,ref30} &
Room temperature; 8~ppm$\cdot$nm \cite{ref22}; 19~ppm$\cdot$nm per orientation (76 total), 9-nm layer \cite{ref30} &
Dipolar XXZ; $-0.50\pm0.13$~dB squeezing \cite{ref22}; Floquet OAT/TAT-like control and 6.7(6)\% amplification \cite{ref30} &
DROID 91~$\mu$s and XY-8 11.5~$\mu$s \cite{ref22}; native-axis XY8 6.5~$\mu$s \cite{ref30} &
Spin squeezing and signal amplification demonstrated with different observables \\
Donor electron spins in $^{28}$Si:P \cite{ref34} &
47~ppm residual $^{29}$Si; $[\mathrm{P}]=5\times10^{11}~\mathrm{cm}^{-3}$ at 4.5~K and $4\times10^{12}~\mathrm{cm}^{-3}$ at 1.7~K &
No collective twisting, squeezing, or amplification benchmark in the cited experiment &
Hahn-echo $T_2 = 130\pm30$~ms and $353\pm9$~ms, respectively &
Long coherence; nonlinear collective-sensing performance not yet established \\
YIG Kittel mode \cite{ref35} &
YIG sphere coupled to a superconducting planar resonator; 10~mK--5~K &
$g_m/(2\pi)=8.17$~MHz is linear magnon-photon coupling; no Kerr or squeezing benchmark &
$\gamma_m/(2\pi)=2.97$~MHz (mode linewidth) &
Linear strong coupling does not by itself quantify nonlinear metrological performance \\
\end{tabular}
\end{ruledtabular}
\end{table*}

NV-center ensembles provide relatively strong dipolar interactions, but their performance is sensitive to spin density and spatial disorder. Random defect positions produce a distribution of dipolar coupling strengths and reduce the uniformity required for collective squeezing~\cite{ref22,ref30}. Donor spins in isotopically enriched $^{28}$Si offer much longer coherence times, with Hahn-echo $T_2$ values reaching hundreds of milliseconds~\cite{ref34}, but collective twisting or squeezing has not yet been demonstrated in the cited experiments. Magnetic systems such as YIG provide strong collective magnon-photon coupling, although the measured linear coupling and magnon linewidth do not directly quantify the nonlinear interaction required for squeezing~\cite{ref35}.

These differences make direct comparison between platforms difficult. Coherence times measured by free induction, Hahn echo, or dynamical decoupling correspond to different experimental conditions, while linear coupling rates cannot be directly compared with nonlinear twisting strengths. Table~\ref{tab:1} summarizes the relevant experimental conditions, nonlinear observables, and coherence or damping scales.

\section{Limitations}

Nonlinear collective-spin sensing is sensitive to disorder, decoherence, and imperfect control. These effects can weaken collective dynamics and reduce quantum enhancement.

Inhomogeneity is an important limitation in solid-state spin ensembles. Variations in local magnetic fields, strain, and spin-spin interactions cause individual spins to evolve differently. This nonuniform evolution reduces coherence and weakens the correlations generated by nonlinear interactions. For dipolar-coupled spins, spatial disorder further produces a broad distribution of interaction strengths.

Control errors provide another limitation. Imperfect microwave or optical pulses can accumulate during long sequences. Magnetic-field and temperature fluctuations can also reduce coherence and degrade the collective state.

Finally, correlation generation must compete with decoherence. Longer nonlinear evolution can generate stronger correlations, but it also increases exposure to noise. Sensing performance therefore typically reaches an optimum at a finite evolution time, after which decoherence reduces the metrological advantage.

\section{Summary}

Nonlinear collective-spin dynamics provide a route to quantum-enhanced sensing by generating correlations among many spins. Quadratic collective interactions produce spin squeezing and other entangled states through nonlinear phase evolution and quantum interference. These states can reduce measurement uncertainty and improve sensitivity beyond the standard quantum limit.

In solid-state platforms, the achievable enhancement depends strongly on the material and experimental conditions. Spin density and interaction strength determine how rapidly correlations develop, while disorder, decoherence, and control imperfections limit how long they remain useful. Experiments with NV-center ensembles demonstrate that these competing effects can be controlled sufficiently to produce spin squeezing and signal amplification, while other platforms offer complementary advantages such as long coherence or strong collective coupling.

Several challenges remain for scalable nonlinear collective-spin sensing. A central goal is to achieve strong and uniform interactions while maintaining long coherence in large spin ensembles. Spatial disorder, surface noise, imperfect control, and inefficient readout can all reduce the metrological gain. Reliable benchmarks are also needed to distinguish genuine many-body enhancement from technical effects and to compare different platforms under realistic experimental conditions.

Future progress will require improvements in both materials and control. Deterministic defect placement, isotopic purification, and interface engineering can reduce disorder and decoherence, while Floquet and cavity-based control can provide more uniform effective interactions. Improved readout and estimation methods can further preserve and extract the generated metrological gain. These developments may enable nonlinear collective-spin sensing to move from demonstrations of squeezing and amplification toward practical quantum-enhanced sensors.

Numerical results presented in this tutorial can be reproduced using the \texttt{tqix} package~\cite{ref36,ref37}.

\subsection*{Recommended / Further reading}

\begin{itemize}
\item[1.] Kitagawa M and Ueda M 1993 Squeezed spin states \textit{Phys. Rev. A} \textbf{47} 5138 \cite{ref17}
\item[2.] Ma J, Wang X, Sun C P and Nori F 2011 Quantum spin squeezing \textit{Phys. Rep.} \textbf{509} 89 \cite{ref18}
\item[3.] Pezz\`e L, Smerzi A, Oberthaler M K, Schmied R and Treutlein P 2018 Quantum metrology with nonclassical states \textit{Rev. Mod. Phys.} \textbf{90} 035005 \cite{ref19}
\item[4.] Degen C L, Reinhard F and Cappellaro P 2017 Quantum sensing \textit{Rev. Mod. Phys.} \textbf{89} 035002 \cite{ref1}
\item[5.] T\'oth G and Apellaniz I 2014 Quantum metrology from a quantum information science perspective \textit{J. Phys. A} \textbf{47} 424006
\end{itemize}

\begin{acknowledgments}
This work is supported by the Tohoku Initiative for Fostering Global Researchers for Interdisciplinary Sciences (TI-FRIS) of MEXT's Strategic Professional Development Program for Young Researchers and FRIS Creative Interdisciplinary Collaboration Program.
\end{acknowledgments}

\bibliography{refs}

\end{document}